\documentstyle[preprint, aps,epsf, floats,eqsecnum]{revtex}

\begin{document}

\preprint{
\vbox{\hbox{JHU--TIPAC-2001-03}
      \hbox{August 2001} }}
      
\title{RS1, Higher Derivatives and Stability}
\author{Adam Lewandowski and Raman Sundrum}
\address{Department of Physics and Astronomy, The Johns Hopkins University \\
  3400 North Charles Street,  Baltimore, Maryland 21218}
\maketitle
\thispagestyle{empty}
\setcounter{page}{0}
\begin{abstract}%
We demonstrate the classical 
 stability of the weak/Planck hierarchy within the
 Randall-Sundrum  scenario,  incorporating the Goldberger-Wise 
mechanism and 
higher-derivative interactions in a systematic perturbative expansion. 
Such higher-derivative interactions are 
expected if the RS model is the low-energy description of some more 
 fundamental theory. Generically, higher derivatives  lead to 
ill-defined singularities in the vicinity of effective field theory branes. 
These are carefully treated by the methods of 
classical renormalization.
\end{abstract}

\newpage

\section{Introduction}

Theories with extra dimensions provide new ways of explaining the weak/Planck
hierarchy. The original proposal for doing so appeared in Ref. \cite{add}. 
An alternative proposal is the Randall-Sundrum (RS1) scenario \cite{rs1},
 where the  hierarchy is set by a relative warp factor, 
$e^{-  k \pi r_c}$, between a ``visible'' brane, to which the Standard 
Model is confined, and a ``hidden'' brane where 4D gravity is highly 
localized by the RS2 mechanism \cite{rs2}. Here, $k$ is a fundamental 
scale determined by the 5D cosmological constant and $r_c$ is the 
compactification radius. The Goldberger-Wise mechanism \cite{gw} provides 
a simple and natural means of stabilizing the radius at  
$k r_c \sim {\cal O}(1/\epsilon)$, by introducing a bulk scalar field 
with 5D mass-squared of order $\epsilon$ in fundamental units. 
The large observed weak/Planck hierarchy, $e^{- k \pi r_c} \sim 10^{-15}$,
is then generated  from a modest fundamental hierarchy, 
$\epsilon \sim 1/10$. 

Since the RS1 field theory, including  general relativity, 
 is quantum-mechanically non-renormalizable, the 
model must be considered to be an effective description of a more 
fundamental theory. Refs.~\cite{rsstring} have discussed  
 string theory embeddings of the RS1 mechanism. In any such 
embedding, higher-derivative interactions ($\alpha'$ corrections in string 
theory) are expected to appear in the effective field theory after 
integrating out very massive physics. It is, 
therefore, important to demonstrate that RS1 and the Goldberger-Wise 
mechanism are stable under the addition of such higher-derivative terms.
In this paper, we will show that this is indeed the case within the 
systematic framework of classical effective field theory. 

While short-distance quantum effects can be parameterized and studied within  
a local derivative expansion, it is also important  to demonstrate 
stability in the presence of genuine, long-distance quantum effects. 
These can also be studied within effective field theory. 
Refs.~\cite{rothstein} have examined 
 such effects at one loop. 
 We hope to give a more complete treatment of quantum effects in future
work.

Recently, a  
dual picture of the RS scenario has been 
developed \cite{adscftrs}, based on the AdS/CFT correspondence \cite{adscft},
 in which the extra-dimensional 
dynamics is replaced by a strongly coupled conformal field theory. 
While this duality is compelling and powerful, we will not make use of it in 
this paper as some aspects remain unproven.

Our strategy is to first set up a systematic perturbative expansion 
for the classical effective field theory in which to study higher derivatives.
While the Goldberger-Wise $\epsilon$ is an obvious small parameter of the 
scenario, we 
cannot perturbatively expand in it since the hierarchy is set by 
$e^{- 1/\epsilon}$, which vanishes to all orders in $\epsilon$. 
Instead, we choose both the bulk curvature and the brane tadpole couplings 
which give the Goldberger-Wise 
scalar a non-trivial profile in the extra dimension to provide our formal 
expansion parameter, $\lambda$. 
In Section II, we re-derive the Goldberger-Wise  
mechanism in the absence of higher derivatives. We note that there 
is an elegant, exactly soluble version of the Goldberger-Wise mechanism 
\cite{freedman}, but our perturbative treatment will be more convenient 
when higher-derivative terms are added. A discussion of the 
Goldberger-Wise mechanism related to ours is Ref. \cite{csaki}.
 In Section III, we discuss how  higher-derivative 
terms are constrained by symmetries. In Section IV we discuss the 
apparent incompatibility of the derivative expansion, normally 
valid at long distances, with the presence  of ``thin'' or 
$\delta$-function branes. We show how the ill-defined singularities that  
arise in the equations of motion can be eliminated by classical 
renormalization. In Section V we demonstrate the stability of the RS 
and Goldberger-Wise mechanisms when higher-derivative perturbations are 
included. The central technical concern  is that terms in our perturbative 
expansion take the form $\lambda^n f_n(\epsilon)$ and it is important 
for a controlled expansion that 
the small parameter $\lambda$ is not overwhelmed by possible large terms in 
$f_n$, such as $e^{1/\epsilon}$ or $1/\epsilon^m$. This is carefully checked.
Section VI provides our conclusions.

\section{The Goldberger-Wise Mechanism in Perturbation Theory}

\subsection{The model}

RS1  has a single extra dimension which is an interval, 
realized as an orbifold, $S^1/{\bf Z}_2$. We will begin by using a 
conventional angular coordinate, $-\pi \leq \phi \leq \pi$, for the 
$S^1$, where the orbifold symmetry acts by $\phi \rightarrow -\phi$. 
We will always describe fields within the fundamental domain $0 \leq \phi 
\leq \pi$. Their extension to general $\phi$ is then determined by 
the orbifold symmetry and periodicity on $S^1$.

The 
Goldberger-Wise mechanism will first be implemented within a theory  given by
\begin{equation}
S= S_{bulk} + 
S_{vis}+S_{hid},
\end{equation}
where
\begin{eqnarray}
S_{bulk} & = &  M^3 \int 
d^4 x \int_{-\pi}^{\pi} d \phi \sqrt{G} \left( - \frac{1}{4}  R + 
\frac{1}{2}(D \chi)^2 + 3 k^2  -
\frac{1}{2} \epsilon k^2 \chi^2 \right) \\
S_{vis} & = & - M^3 k \int 
d^4 x \sqrt{g_{v}} \left(\rho_{v} + \lambda_{v} \chi + \frac{1}{2} \mu_v 
\chi^2 \right) \\
S_{hid} & = & - M^3 k \int d^4 x \sqrt{g_{h}} 
\left(\rho_{h} + \lambda_{h} \chi + \frac{1}{2} \mu_h \chi^2 \right),
\end{eqnarray}
with
\begin{eqnarray}
g^{v}_{\mu \nu}(x) &\equiv& G_{\mu 
\nu}(x, \phi = \pi) \nonumber \\
g^{h}_{\mu \nu}(x) &\equiv& G_{\mu \nu}(x, 
\phi = 0).
\end{eqnarray}
Note that we have chosen a normalization such that 
$\chi$ is dimensionless. We assume that there are no extremely large 
hierarchies among the couplings of the model. However, we do take 
\begin{eqnarray}~\label{epsiloncondition}
k < M \nonumber \\
|\lambda_{v,h}| < \epsilon < |\rho_{v,h}| 
\sim \mu_{v,h} \sim 1. 
\end{eqnarray}

We will restrict our attention to 
classical solutions which admit a four-dimensional
Poincare invariance. Such configurations satisfy the ansatz
\begin{eqnarray}
ds^2 & = & e^{2 A(\phi)} \eta_{\mu \nu} dx^{\mu} dx^{\nu} - 
r_c^2 d \phi^2 \nonumber \\
\chi & = & \chi(\phi),
\end{eqnarray}
where $r_c$ is the (constant in this ansatz) ``radius''.
In solving the equations of motion it  is convenient to work with a 
re-scaled dimensionless extra-dimensional coordinate, 
\begin{eqnarray}
y &\equiv& k r_c \phi,  
\end{eqnarray}
so that the infinitesimal distance in the extra-dimension is $dy/k$.
The Poincare ansatz then reads, 
\begin{eqnarray}
ds^2 & = & e^{2 A(y)} \eta_{\mu \nu} dx^{\mu} dx^{\nu} - dy^2/k^2 \nonumber \\
\chi & = & \chi(y).
\end{eqnarray}
The equations of motion subject to this ansatz are,
\begin{eqnarray}
\chi'' + 4 A' \chi' - \epsilon \chi & = & (\lambda_{v} +\mu_v \chi) 
\delta (y - y_c) + (\lambda_{h} + \mu_h \chi) \delta (y) \label{a} \\
A'' + \frac{2}{3} (\chi')^2 & = & -\frac{2}{3} (\rho_{v} + \lambda_{v} 
\chi + \frac{1}{2} \mu_v \chi^2 ) \delta(y-y_c) \nonumber \\
& & \mbox{} - 
\frac{2}{3} (\rho_h +  \lambda_{h} \chi +\frac{1}{2} \mu_h \chi^2 ) 
\delta(y) \label{b} \\
(A')^2 & = & 1 - \frac{1}{6} \epsilon \chi^2 + \frac{1}{6} (\chi')^2, 
\label{c}
\end{eqnarray}
where 
\begin{equation}
y_c \equiv k r_c \pi.
\end{equation}

The brane-localized scalar tadpoles provide us with our formal small 
expansion parameters, $\lambda_{v,h} \sim {\cal O}(\lambda)$.  
We expand our solution as a perturbation series in $\lambda$,  
\begin{eqnarray}~\label{expansion}
A & = & \sum_n A_n \nonumber \\
\chi & = & \sum_n \chi_n,
\end{eqnarray}
where the subscript $n$ denotes the term of order $\lambda^n$ in the series.
Notice that at order $\lambda^{0}$ we have $\chi=0$ and the (truncated)
$AdS_5$ gravity solution obtained in RS1, $A_0 = -y$, with $r_c$ arbitrary.
Thus, in fluctuations away from 4D Poincare invariance, $r_c$ becomes a 
``radion'' modulus at zeroth order. 
We will demonstrate the Goldberger-Wise mechanism for stabilizing the radius 
in higher orders of perturbation theory. 

The strategy for solving the equations of motion is as follows.  
We solve for $A'$ in terms of $\chi$ using (\ref{c}),
\begin{equation}~\label{A}
A' =  - \left[1 - \frac{1}{6} \epsilon \chi^2 + \frac{1}{6} (\chi')^2 \right]^{1/2}, 
\end{equation}
 and eliminate $A'$ from (\ref{a}) to obtain an equation purely for $\chi$, 
\begin{equation}~\label{chi}
 \chi'' - 4 \chi' \sqrt{1 - \frac{1}{6} \epsilon 
\chi^2 + \frac{1}{6} (\chi')^2}  - \epsilon \chi  =  
(\lambda_{v} + \mu_v \chi) \delta (y - y_c) + (\lambda_{h} 
+ \mu_h \chi) \delta (y).
\end{equation}
We will solve (\ref{chi}) to any desired order in $\lambda$. We then 
integrate 
(\ref{A}) to solve for $A(y)$, subject to the canonical gauge choice 
$A(0) = 0$. 
Equation (\ref{b}) will then 
be automatically solved away from the branes as a consequence of 5D general 
covariance.\footnote{One can easily check that up to $\delta$-function terms, 
(\ref{a}) and (\ref{c}) imply (\ref{b}).}  
Finally, we satisfy the two brane junction conditions of (\ref{b}) by 
fine-tuning the hidden brane tension parameter $\rho_h$ (equivalent to 
fine-tuning the  4D cosmological constant to zero), and adjusting the 
compactification radius, $r_c$ (or equivalently, $y_c$) to its 
stable vacuum value.

\subsection{Perturbation theory}

Here, we will show self-consistently that the solution of the equations 
of motion satisfies,
\begin{eqnarray}~\label{claim}
y_c &\sim& {\cal O}(1/\epsilon) \nonumber \\
A'(y) &\approx& - 1,
\end{eqnarray}
so that  the weak/Planck hierarchy determined by the RS1 mechanism is set 
by $e^{-{\cal O}(1/\epsilon)}$. Therefore, while we need $\epsilon$ to be 
somewhat small, we cannot work perturbatively 
in $\epsilon$. (As stated above we will be working strictly perturbatively 
in $\lambda$.) However, we will simplify our analysis by dropping 
subleading terms in  $e^{-{\cal O}(1/\epsilon)}$. 
Note that while $\lambda$ is formally small, combinations such as 
$\lambda {\cal O}(1/\epsilon^m)$ or $\lambda {\cal O}(e^{+ 1/\epsilon})$
may be large when $\epsilon \ll 1$. 
It is, therefore, crucial that such combinations do not appear in 
the perturbative series in order for the perturbative expansion to be
under control. We show that this
danger does not eventuate by a careful analysis.

\subsubsection{Stabilization at second order}

Since $\chi$ vanishes at zeroth order, we can perturbatively expand the 
square-root in (\ref{chi}). 
  At first order,
\begin{equation}
\chi''_1 - 4 \chi'_1 - \epsilon \chi_1 = (\lambda_v + \mu_v \chi_1) 
\delta(y-y_c) + (\lambda_h + \mu_h \chi_1) \delta(y).
\end{equation}
With  orbifold boundary conditions the solution is 
\begin{equation}~\label{chi1}
\chi_1 = c_1 e^{\Delta_+ (y-y_c)} + c_2 e^{\Delta_- y},
\end{equation}
with
\begin{eqnarray}
c_1 & \approx & \frac{-\lambda_v}{(2 \Delta_+ + \mu_v)}
-\frac{\lambda_h (2 \Delta_-+ \mu_v) 
e^{\Delta_- y_c}}{(2 \Delta_+ + \mu_v)(2 \Delta_- - \mu_h)} \\
c_2 & \approx & \frac{\lambda_h}{(2 \Delta_- - \mu_h)},
\end{eqnarray}
and where 
\begin{equation}
\Delta_{\pm} = 2 \pm \sqrt{4+\epsilon}.
\end{equation}
Note that for small $\epsilon$, 
\begin{equation}~\label{delta}
\Delta_+ \approx 4, ~~\Delta_- \approx - 
\epsilon/4.
\end{equation} 
In the expressions for the coefficients we have neglected subleading 
powers of $e^{-\Delta_+ y_c}$, since we will demonstrate (\ref{claim}).

At first order in $\lambda$ the scalar field does not stabilize the radion 
because $A$ receives no correction at this order.  
There are still two fine tunings  needed to satisfy the equations of motion
as in the original version of RS1 without 
stabilization. 
 It is only at second order, where the  first order 
scalar profile back-reacts on the metric, that the   
compactification radius is fixed. This back-reaction was also 
discussed in Ref. \cite{csaki}. 
%This is essentially equivalent to the minimization of the effective 
%4D potential in [?].  
 At second order $\chi_2 = 0$ and $A'_2$ is given by (\ref{c}) expanded to 
second order in $\chi$.  The junction conditions of (\ref{b}) at this order 
read,
\begin{eqnarray}
\left( -1 + \frac{1}{12} \epsilon \chi_1(0)^2-\frac{1}{12} 
\chi'_1(0)^2 \right) & = & -\frac{1}{3} \left( \rho_h+\lambda_h 
\chi_1(0) + \frac{1}{2} \mu_h \chi_1 (0)^2\right) \\
 \left( -1 + \frac{1}{12} \epsilon \chi_1(y_c)^2-\frac{1}{12} 
\chi_1'(y_c)^2 \right) & = & \frac{1}{3} \left( \rho_v+\lambda_v 
\chi_1(y_c) + \frac{1}{2} \mu_v \chi_1(y_c)^2 \right). 
\end{eqnarray}

Note that parametrically in $\lambda$, the only way the visible junction 
condition can be solved is if 
\begin{eqnarray}
% \rho_h &=& 3 + {\cal O}(\lambda^2) \nonumber \\
\delta \rho_v &\equiv&  3 + \rho_v  \sim {\cal O}(\lambda^2).
\end{eqnarray}
Such a condition will not reappear at higher orders in $\lambda$.  
Once we grant that $\rho_v$ is somewhere 
in this ${\cal O}(\lambda^2)$-sized window about $-3$, the visible junction 
 condition can be satisfied,  not by fine tuning of couplings, 
but by solving for the stable vacuum value of the dynamical radius, 
\begin{equation}~\label{yc}
y_c = \frac{\ln (\Sigma)}{ \Delta_-},
\end{equation}
where
\begin{eqnarray}
&\Sigma & =  \frac{ \Delta_+ (2 \Delta_- - \mu_h)(2 \Delta_- -2-\mu_v) 
\lambda_v}{(\Delta_+-\Delta_-)(4 \Delta_- \Delta_+ + \mu_v (2+ \mu_v)) 
\lambda_h} \nonumber \\
 & & \mbox{} \pm \frac{(\mu_h-2 \Delta_-)(\mu_v + 2 \Delta_+) 
\sqrt{(2 \delta \rho_v (4 \Delta_- \Delta_+ + \mu_v(2+\mu_v)) 
+ (1-\Delta_- \Delta_+ + \frac{\mu_v}{2}) 
\lambda_v^2)}}{(\Delta_+-\Delta_-)(4 \Delta_- \Delta_+ + 
\mu_v (2+ \mu_v)) \lambda_h}.
\end{eqnarray}
The sign will depend on the actual value of the parameters with the 
requirement that $y_c$ is real and positive. It follows that in fluctuations
away from 4D Poincare invariance, the associated radion has acquired a 
mass-squared at this order in $\lambda$. For a large range of the parameters,
this mass-squared is positive. For example, if $\mu_{v,h}$ dominate over 
$\Delta_{\pm}$ in (\ref{chi}), then in this limit the computation of the 
radion effective potential at second order is 
precisely the one performed by Goldberger and Wise, 
 with the identification of their 
$v_{v,h}$ with our $-\lambda_{v,h}/\mu_{v.h}$. A positive mass-squared 
results \cite{gw,gw2,csaki}.

The hidden junction condition at $y=0$ 
gives us a fine tuning condition for $\rho_h$.
\begin{equation}
\rho_h = 3+\frac{(2\Delta_- (-2 + \Delta_- \Delta_+) + 
\mu_h) \lambda_h^2}{4 (\mu_h-2 \Delta_-)}.
\end{equation}
This  fine-tuning which we must do is equivalent to fine-tuning the 
effective four-dimensional cosmological constant to zero in order to permit 
solutions with 4D Poincare invariance. We must perform such a fine-tuning
order by order in $\lambda$. 

Let us now check that our basic claims (\ref{claim}) are satisfied at this 
order. For generic values of the couplings, 
\begin{equation}
{\rm ln} \Sigma \sim {\cal O}(1),
\end{equation}
so by (\ref{delta}), $y_c \sim {\cal O}(1/\epsilon).$
Given the explicit form for $\chi_1$ it is straightforward to see that 
the resulting $A_2$ satisfies, $A' \approx -1$. 
We will show that these successes are maintained at higher order in $\lambda$.

\subsubsection{Subdominance of higher orders}

At higher order $n \geq 2$, after expanding the square-root in (\ref{chi}), 
 we must solve the following equation for $\chi$
\begin{equation}~\label{recurrence}
\chi''_n  - 4  \chi'_n - \epsilon \chi_n -\mu_v \chi_n \delta(y-y_c) - 
\mu_h \chi_n \delta(y)  =  J_n,
\end{equation}
where $J_n$ has  the form 
\begin{equation}~\label{rootexp}
J_{n} =   \sum a_{r, s, \vec{i}} 
\epsilon^{r} \chi_{i_1} \ldots \chi_{i_{2r}} 
\chi'_{i_{2r+1}} \ldots \chi'_{i_{2(s+r)+1}} 
\delta_{i_1 + \cdots + i_{2(s+r) + 1},~n}.
\end{equation}
Here, the expansion coefficients, $a_{r, s, \vec{i}}$, are numbers of 
order one completely 
determined by expanding the square-root in (\ref{chi}), and $r+s \geq 1$.
Note that $J_n$ is determined from lower order solutions of $\chi$
 in perturbation theory.  
This allows us to treat $J_n$ as a known source in the 
$n^{th}$ order equation 
of motion for $\chi$. Thus we can solve (\ref{recurrence}) iteratively 
for the $\chi_n$.

Our central
 claim is that to any order in perturbation theory $\chi$ has the form, 
\begin{equation}~\label{chin}
\chi_n = \sum b_{\vec{n}, r, s, t} \lambda^{n_1} (\lambda y)^{n_2} 
(\lambda y_c)^{n_3} 
(\lambda/\Delta_-)^{n_4} 
e^{r \Delta_+ (y - y_c)} e^{s \Delta_- y} e^{t \Delta_- y_c}, 
\end{equation}
where the constant 
coefficients $b$ are order one or smaller (they can contain 
positive but not negative powers of $\epsilon$), and $n_i, r, s, t$ are 
integers such that 
\begin{eqnarray}
n_1 &\geq& 1; ~~ r, n_2, n_3, n_4 \geq 0 \nonumber \\
n &=& n_1 + n_2 + n_3 + n_4 \nonumber \\
|s|, |t| &\leq& n. 
\end{eqnarray} 
Let us discuss the significance of this claim.
First, it
 shows that the perturbative expansion is well-behaved. Note that this is 
{\it a priori} not guaranteed. For example, the formal small parameter $\lambda$ 
could be overwhelmed by a factor of $e^{1/\epsilon}$ or a high enough 
power of 
$1/\epsilon$. However, it is straightforward to see that this is not so given 
(\ref{chin}).  Although 
$y \leq y_c \sim {\cal O}(1/\Delta_-) \sim {\cal O}(1/\epsilon)$, 
powers of $y$, $y_c$, and 
$1/\Delta_-$ are accompanied by powers of the 
parametrically smaller $\lambda$. While positive 
powers of $e^{\Delta_+ y} > 1$ 
appear, they are compensated by powers of $e^{- \Delta_+ y_c}$. While 
arbitrary powers of $e^{\Delta_- y}$ can appear, since we will 
show (\ref{claim}) self-consistently,  $e^{\Delta_- y} \sim {\cal O}(1)$. 
Secondly, (\ref{claim}) also follows since it is clear from (\ref{chin}) 
that $\chi$ is dominated by $\chi_1$ for all $y$. Therefore, the second 
order determination we made for $A'$ dominates, and leads to a stable 
value of $y_c$ near that of (\ref{yc}). The hidden brane 
junction condition also receives subleading corrections which can be 
satisfied by fine-tuning $\rho_h$ order by order in $\lambda$. This is 
the higher-orders incarnation of the four-dimensional cosmological constant
fine-tuning problem of ensuring a 4D Poincare invariant vacuum 
solution.

We will now prove our claim (\ref{chin}) for $\chi_n$
by induction in $n$. First we note that 
the claim is true for $n = 1$ based on our explicit solution. For a general 
perturbative order, $n$, we assume that the claim is true for all lower 
orders. Then clearly $J_n$, constructed from the lower order solution for 
$\chi$, also has the form
\begin{equation}~\label{Jn}
J_n = \sum c_{\vec{n}, r, s, t} \lambda^{n_1} (\lambda y)^{n_2} 
(\lambda y_c)^{n_3} 
(\lambda/\Delta_-)^{n_4} 
e^{r \Delta_+ (y - y_c)} e^{s \Delta_- y} e^{t \Delta_- y_c}, 
\end{equation}
where the constant coefficients, $c$, are also order one or smaller.
Given this source term we can solve (\ref{recurrence}), 
\begin{equation}~\label{convolution}
\chi_n = \int_0^{y_c} G(y, y') J_n(y') dy',
\end{equation}
where $G$ is the Green function satisfying
\begin{equation}~\label{Gdef}
\left\{ \frac{d^2}{d y^2} - 4 \frac{d}{d y}  - \epsilon  - \mu_v  
\delta(y-y_c) - \mu_h  \delta(y) \right\} G(y, y') = \delta(y-y'),
\end{equation}
subject to orbifold boundary conditions. 

The detailed form of the Green function is straightforwardly worked out. 
However, 
in order to complete our induction we do not require the full details, but 
only the general form,
\begin{equation} ~\label{G}
G(y,y') = \left\{  \begin{array}{lr} \left({\cal O}(1) e^{- \Delta_+ y'}+ 
{\cal O}(1)  e^{\Delta_- (y_c-y') - \Delta_+ y_c} \right) e^{\Delta_+ y} + & \\ 
\qquad \left({\cal O}(1) e^{- \Delta_+ y'}+ {\cal O}(1)  e^{\Delta_- (y_c-y') 
- \Delta_+ y_c} \right) e^{\Delta_- y}, &  0<y<y' \\
\left( {\cal O}(1) e^{\Delta_- (y_c-y') - \Delta_+ y_c} + {\cal O}(1) 
e^{\Delta_- y_c - \Delta_+ (y'+ y_c)}\right) e^{\Delta_+ y} + & \\ \qquad 
\left({\cal O}(1) e^{-\Delta_- y'}+{\cal O}(1) e^{-\Delta_+ y'}\right)e^{\Delta_- y}, 
&  y'<y<y_c.  \end{array} \right.
\end{equation}
The proof that $\chi_n$ satisfies the claim of (\ref{chin}) then follows 
by inspection of all the possible terms that can arise from 
(\ref{convolution}) given (\ref{Jn}) and (\ref{G}). The basic integrals 
involved in evaluating (\ref{convolution}) are of the form,
\begin{equation}~\label{int1}
\int_{y_1}^{y_2} dy' y'^m = \frac{y_2^{m+1} - y_1^{m+1}}{m+1}, 
\end{equation}
or
\begin{equation}~\label{int2}
\int_{y_1}^{y_2} dy' y'^m e^{\Delta y'} = 
\frac{d^m}{d \Delta^m} \left(
\frac{e^{\Delta y_2} - e^{\Delta y_1}}{\Delta} \right),
\end{equation}
where $m$ is a non-negative integer, $\Delta$ is some linear combination
of $\Delta_{\pm}$ with integer coefficients, 
and $y_1 = 0, y; ~ y_2 = y, y_c$. It is straightforward 
but a little tedious to then check that all possible terms arising in 
(\ref{convolution}) do indeed satisfy the claim of (\ref{chin}).

\subsubsection{The case of $\lambda > \epsilon$    }

We conclude this reworking of the Goldberger-Wise mechanism by discussing 
the relationship between $\lambda$ and $\epsilon$. Thus far, we have 
considered $\lambda$, our formal perturbative expansion parameter, to be 
formally smaller than $\epsilon$. The reason for doing this is because
of terms in (\ref{chin}) with non-zero $n_2, n_3$ or $n_4$, all of which 
would signify effects of order $(\lambda/\epsilon)^{n_i}$. The presence of
 such terms threatens to invalidate perturbation theory if we allowed 
$\lambda > \epsilon$. However, we can in fact prove a stronger result, that 
all such terms in (\ref{chin}), are in fact accompanied by a suppression 
factor of order $\epsilon^{n_2 + n_3 + n_4}$, so that we can take 
take $\lambda > \epsilon$ (though formally small compared to unity). 
This relationship is indeed implicit in the 
original analysis of Goldberger and Wise. We have not 
emphasized this fact until now both for simplicity and 
because the suppression factor is not generally true
when we study higher derivative perturbations. 
We will study the special circumstances under which higher-derivative 
perturbations  do not require $\lambda/\epsilon$ to be small 
in Section V.  

The stronger result, that there are accompanying factors of 
order $\epsilon^{n_2 + n_3 + n_4}$, is 
again proven by induction on $n$, noting first that it trivially 
holds for $n= 1$. For a 
larger $n$, we assume it holds for lower orders. Therefore, $J_n$ shares 
the same property since it is made from products of (lower-order) $\chi$ 
and $\chi'$. But $\chi_n$ determined by (\ref{convolution}) can generate 
(at most) 
one new power of $y, y_c$ or $1/\Delta_-$ relative to $J_n$: a new power 
of $y$ or $y_c$ can arise as in (\ref{int1}) or (\ref{int2}), 
or one new power of 
$1/\Delta_-$ can arise as in (\ref{int2}) in case $\Delta = \Delta_-$.
One can check (somewhat tediously) that such cases can only arise from 
terms in $J_n$ involving at least one power of $\chi$ without a derivative 
acting on it ($r>1$ in (\ref{rootexp})).
 Otherwise, in terms where the integrals (\ref{int1}, \ref{int2}) produce 
an extra power of $y, y_c$ or $1/\Delta_-$, the derivatives in $\chi'$
always bring down a power of $
\Delta_- \sim {\cal O}(\epsilon)$ or eliminate a power of $y'$ 
so that in fact $\chi_n$ does satisfy the claim of the induction. 
But for terms in $J_n$ with at least one non-derivative power of $\chi$,
we see in (\ref{rootexp}) that there are explicit powers of 
$\epsilon$ arising in the 
expansion (\ref{rootexp}). These compensate the single power of 
of ${\cal O}(1/\epsilon)$ that can be generated in $\chi_n$, so the 
induction claim still goes through.

\section{Symmetries of interactions}

Our task now is to specify which 
 higher-derivative interactions are allowed in generalizing RS1 and the 
Goldberger-Wise mechanism, in particular how they are constrained by 
symmetries. Since the spacetime manifold is topologically 
${\bf R}^4 \times S^1/{\bf Z}_2$, the relevant symmetries are 5D general 
coordinate invariance and the ${\bf Z}_2$ parity symmetry. 
The coordinate-invariant or geometric statement of the ${\bf Z}_2$ symmetry 
is as follows. We first start with the manifold ${\bf R}^4 \times S^1$ and 
consider two ``3-branes'' which divide the $S^1$ into two disjoint 
regions. We choose geometries and scalar fields on ${\bf R}^4 \times S^1$ 
such that the two regions are reflection symmetric. 
We now wish to find a convenient 
description of all interactions which respect this. 

It is useful to start with a formalism  which respects full 5D general 
coordinate invariance on ${\bf R}^4 \times S^1$, even though this 
necessarily involves coordinate systems which do not respect the 
${\bf Z}_2$-symmetry of the invariant geometry. (However, see 
Ref.~\cite{rattazzi} for an alternative symmetry implementation.)
 The two fixed-point 
branes (``hidden'' and ``visible'') will be coordinatized as 
$Y^M_{hid}(x)$, $Y^M_{vis}(x)$, where $x$ are parameterizations of the
 3-branes. 
Using these brane fields, a 5D metric, $G_{MN}(X)$, and 5D 
Goldberger-Wise scalar $\chi(X)$, we can 
form fully 5D general coordinate-invariant bulk and brane actions on 
${\bf R}^4 \times S^1$, as in Ref. \cite{braneeft}.

%We must further restrict such actions 
%to be invariant under the reflection symmetry, so that the associated 
% dynamics is compatible with orbifolding.

%Obviously, 
%such actions can  automatically be 
%restricted to give actions on ${\bf Z}_2$-symmetric 
%invariant geometries and scalar fields. 

Once a generally coordinate invariant  action has been chosen in this way, 
 we can 
 ``gauge-fix'' by choosing our 5D coordinates, $X^M$, to consist of 
$x^{\mu}$ and an extra-dimensional angle, $-\pi \leq \phi \leq \pi$, 
such that, 
\begin{eqnarray}~\label{gaugefix}
Y_{hid}^{\mu}(x) &\equiv& Y_{vis}^{\mu}(x) \equiv x^{\mu} \nonumber \\
Y_{hid}^{\phi} &\equiv& 0 \nonumber \\
Y_{vis}^{\phi} &\equiv& \pi,
\end{eqnarray}
and where the symmetry of the geometry and $\chi$ is manifest, 
\begin{eqnarray}~\label{gaugefix2}
G_{\mu \nu}(x, \phi) &=& G_{\mu \nu}(x, - \phi) \nonumber \\
G_{\mu \phi}(x, \phi) &=& - G_{\mu \phi}(x, - \phi) \nonumber \\
G_{\phi \phi}(x, \phi) &=&  G_{\phi \phi}(x, - \phi) \nonumber \\
\chi(x, \phi) &=& \chi(x, - \phi).
\end{eqnarray}

Our procedure for writing actions on ${\bf R}^4 \times S^1/{\bf Z}_2$ is 
therefore to (a) write a general 5D coordinate invariant action 
for bulk and brane fields 
on ${\bf R}^4 \times S^1$ as detailed in Ref.~\cite{braneeft},  (b) 
gauge-fix the action according to (\ref{gaugefix}) and (\ref{gaugefix2}), 
using the 
${\bf Z}_2$-symmetry of the allowed configurations. Once this is done 
the brane-fields $Y^M_{hid}(x)$, $Y^M_{vis}(x)$, no longer explicitly appear. 
Finally, (c) one must check that all terms in the action are compatible with 
orbifolding, namely they are invariant under 
\begin{eqnarray}
G_{\mu \nu}(x, \phi) &\rightarrow& G_{\mu \nu}(x, - \phi) \nonumber \\
G_{\mu \phi}(x, \phi) &\rightarrow& - G_{\mu \phi}(x, - \phi) \nonumber \\
G_{\phi \phi}(x, \phi) &\rightarrow&  G_{\phi \phi}(x, - \phi) \nonumber \\
\chi(x, \phi) &\rightarrow& \chi(x, - \phi),
\end{eqnarray}
for general metrics and scalar field.

It is straightforward to check that it does not matter whether one imposes 
the ${\bf Z}_2$-symmetry and gauge-fixing on
the action or varies the un-gauge-fixed action without imposing the 
${\bf Z}_2$-symmetry, as long as one then 
imposes the ${\bf Z}_2$ and  gauge-fixing
 on the equations of motion. For convenience we will assume that the 
${\bf Z}_2$-symmetry and gauge-fixing have been imposed at the level of the 
action, as implicit in the RS1 and Goldberger-Wise papers. 

Recall that 
the central reason we wish to orbifold is that 
we wish to take the visible brane to have ``negative tension'', which would 
normally yield a ghost-like sign for the associated $Y_{vis}^{\phi}$ 
kinetic term (Ref.~\cite{braneeft}), 
causing a vacuum instability towards violent 
brane fluctuations. However, such brane fluctuations violate the 
${\bf Z}_2$-symmetry, and therefore are eliminated by orbifolding. 
It may seem that by formally reintroducing $Y_{vis}^{\phi}$ in step (a) 
of our procedure for generating allowed action terms, we are reintroducing 
the instability. However this is not so. The ${\bf Z}_2$-symmetry 
renders  $Y_{vis}^{\phi}$ as pure gauge, with no physical 
import, as reflected in the gauge fixing of step (b).

\section{Higher-Derivative Operators in Effective Brane Theories}

In any effective field theory we expect there to be higher-derivative 
operators which are remnants of the more fundamental physics which has been 
integrated out. Their interpretation and treatment  is generally 
well-understood. However, 
they pose special problems when they appear in effective theories involving 
orbifolds and 
branes in extra dimensions. These stem from their appearance 
 as $\delta$-function sources in the extra dimension, rather
than some structure with a finite size (perhaps due to quantum-mechanical or 
stringy effects). High derivatives applied to such $\delta$-functions 
produce messy and ill-defined equations of motion. We will show how such 
problems can be treated 
by a combination of field redefinitions and classical brane 
renormalization. See Ref.~\cite{wise} for a recent discussion of 
brane renormalization focusing on co-dimension 2, and Refs.~\cite{damour} 
for earlier work on classical renormalization.

\subsection{The formal derivative expansion}

To keep only the essential considerations in focus we will not consider 
gravity here. We will limit ourselves 
to a single scalar field, $\chi$, living on an extra-dimensional $S^1/{\bf Z}_2$ 
of fixed radius $r_c$ (and the usual four dimensions). The same 
considerations apply in the presence of gravity or non-trivial warp-factor. 
As before we take $\chi$ to be dimensionless and even under the orbifold 
parity. The general Lagrangian then has the form,
\begin{equation}~\label{general}
{\cal L} = M^3 \left\{ \frac{K(\chi)}{2} (\partial_M \chi)^2 - V(\chi) 
- v_i(\chi) \delta (y - y_i) + \sum_{n>0} \frac{1}{M^n} 
{\cal L}_{h.d.}^{(n)} (\chi, \partial_M) \right\},
\end{equation}
where $M$ is the only explicit scale appearing in order to balance dimensions,
and 
$y$ is now defined as
 the extra-dimensional coordinate corresponding to proper distance 
in the extra dimension. Brane localized terms are represented by 
multiplying by one of $\delta(y - y_i)$, where $y_1 = 0, y_2 = \pi r_c$.
${\cal L}_{h.d.}$ are arbitrary higher-derivative terms organized in powers 
of $1/M$. Note that derivatives always appear in even numbers due to the 
orbifold parity and 4D Lorentz invariance, so that the $1/M$-suppressed terms 
(once the overall $M^3$ is excluded) are those with more than two derivatives
in the bulk and those with any deriviatves on a brane. We have not separated 
out brane and bulk higher-derivative terms although we will do this later.

We will study the equations of motion subject to the 4D Poincare ansatz, 
$\chi = \chi(y)$, 
\begin{equation}~\label{generaleom}
- \partial_y (K(\chi) \partial_y \chi) - \frac{K'(\chi)}{2} 
(\partial_y \chi)^2 + V'(\chi) + v_i'(\chi) \delta(y - y_i) = \sum_{n>0} 
\frac{1}{M^n} \frac{\delta S^{(n)}_{h.d.}}{\delta \chi}.
\end{equation}
This would appear to be a differential equation of arbitrarily high order 
and therefore requiring an arbitrary number of initial conditions to solve. 
However, there is a unique solution which 
is perturbatively close, in a $(\partial/M)^m$ expansion, to the zeroth order 
solution,
\begin{equation}
- \partial_y (K(\chi_0) \partial_y \chi_0) - \frac{K'(\chi_0)}{2} 
(\partial_y \chi_0)^2 + V'(\chi_0) + v_i'(\chi_0) \delta(y - y_i) = 
0.
\end{equation}
This zeroth order equation is manifestly a second order differential equation 
subject to orbifold boundary conditions, which we know has a distinct 
solution. Of course, in order to determine $\chi_0$ we may have to approximate 
as we did for the Goldberger-Wise mechanism, perturbing in $\lambda$, but 
in order to focus on higher derivative perturbations let us take the 
zeroth order solution, $\chi_0$, as given. 

An expansion in $\partial_y/M$ 
may seem sensible in the bulk where solutions are smooth
 but ill-defined in the presence of the brane $\delta$-functions. We will 
 show how these ill-defined brane singularities can be renormalized away.
To get started and to understand the issues 
we will proceed bravely and formally define the 
$\partial_y/M$ perturbation expansion.\footnote{This is in the same spirit 
as setting up the formal Feynman diagram series in quantum field theory, 
which is also ill-defined until after regularization and renormalization.}
We will work inductively in $n$. Let the expansion of $\chi$ in powers of 
$1/M$ be written, 
\begin{equation}~\label{fred}
\chi = \sum_m \chi_m.
\end{equation}
Suppose that we can solve (\ref{generaleom}) 
to order $1/M^m$ (we know we can do this for $m=0$). 
We will then show how to construct 
a solution up to order $1/M^{m+1}$.

We substitute (\ref{fred}) into (\ref{generaleom}) 
and focus on  the term precisely 
of order $1/M^{m+1}$, 
\begin{eqnarray}
\Big\{ 
 -\partial_y K(\chi_0) \partial_y - \partial_y (\partial_y \chi_0) K'(\chi_0)
 + \frac{1}{2} K''(\chi_0)(\partial_y \chi_0)^2 & &  \nonumber \\
 + K'(\chi_0)   
(\partial_y \chi_0) \partial_y + 
V''(\chi_0) + v_i''(\chi_0) \delta(y - y_i) 
\Big\}  \chi_{m+1} & = & \frac{\delta S}{\delta \chi} 
\left[ \sum_{\ell \leq m} \chi_{\ell}(y) \right] \bigg|_{m+1}, 
\end{eqnarray} 
where the right-hand side is the functional derivative of the {\it entire} 
action, including the higher derivative terms, evaluated for
 the field $\sum_{\ell \leq m} \chi_{\ell}(y)$, 
but keeping precisely the terms of 
order  $1/M^{m+1}$. 
Thus, in order to perturbatively improve our solution by one order in $1/M$,
that is solve for $\chi_{m+1}$ in terms of $\chi_{\ell \leq m}$, 
we only need to solve 
the above linear second order equation for $\chi_{m+1}$ subject to 
orbifold boundary conditions, with a source term determined by the lower 
order solution $\chi_{\ell \leq m}$. This we can do,  
\begin{equation}~\label{genconv}
\chi_{m+1}(y) = \int dy' G(y, y') 
\frac{\delta S}{\delta \chi} \left[ \sum_{\ell \leq m} \chi_{\ell}(y') \right] \bigg|_{m+1},
\end{equation}
where $G$ is the Green function determined by orbifold boundary conditions 
satisfying,
\begin{eqnarray}
\Big\{ -\partial_y K(\chi_0) \partial_y - \partial_y (\partial_y \chi_0) 
K'(\chi_0)
 + \frac{1}{2} K''(\chi_0)(\partial_y \chi_0)^2 \qquad  & &  \nonumber \\
  + K'(\chi_0)   
(\partial_y \chi_0) \partial_y + V''(\chi_0) + v_i''(\chi_0) \delta(y - y_i) 
\Big\} G(y,y')  & = & \delta (y-y').
\end{eqnarray}

\subsection{The problem}

Such  Green functions, $G$, 
(and also 
 $\chi_0$) will clearly be smooth  in the
bulk, with 
absolute-value 
type kinks at the branes. 
That is, their first derivatives will 
have step-function discontinuities on the branes and their second
derivatives will have $\delta$-functions on the branes.
Let us suppose this property of the Green function is shared by 
$\chi_{\ell \leq m}(y)$  and see 
how trouble can arise in $\chi_{m+1}$.
 First, consider bulk terms that can appear in 
$\frac{\delta S}{\delta \chi} 
[\sum_{\ell \leq m} \chi_{\ell}(y')]_{|_{m+1}}$. If there are 
more than two derivatives acting on the same field, then we will have 
derivatives of $\delta$-functions on the branes which we must convolute 
with $G$, which has absolute-value type kinks. The result is therefore 
ill-defined. Similarly, if we have a power greater than one of second 
derivatives of fields we will have to integrate products of $\delta$-functions
which is again ill-defined.
On the other hand, if there is at most a single field with 
two derivatives acting on it, multiplied by any number of fields with at most 
first derivatives, then we will only have to integrate a single 
$\delta$-function multiplied by a function with absolute-value type 
kinks.\footnote{The first derivatives of fields actually have 
step-function discontinuities, but because of the orbifold symmetries they 
must always appear in even powers, yielding a function with only 
absolute-value type kinks on the branes.}
This does yield a  well-defined $\chi_{m+1}$, which is also smooth except 
for absolute-value type kinks on the branes. It is rather straightforward 
to see that the troublesome higher-derivative bulk terms in the equations 
of motion correspond precisely to bulk higher-derivative terms in 
the Lagrangian, (\ref{general}), with more than one derivative acting on 
a field (unless it can be eliminated by integration by parts). 

Let us now turn to brane terms on the right-hand side of (\ref{genconv}), 
again assuming that $\chi_{\ell \leq m}(y)$ 
has only absolute-value type kinks and 
considering the effects on $\chi_{m+1}(y)$. It is clear that if in the 
Lagrangian, (\ref{general}), there are any derivatives in brane terms, then 
in $\frac{\delta S}{\delta \chi} 
[\sum_{\ell \leq m} \chi_{\ell}(y')]_{|_{m+1}}$ we will have 
derivatives of $\delta$ functions which makes (\ref{genconv}) ill-defined 
again. On the other hand if there are no derivatives on the 
brane terms,
they will give simple $\delta$-functions to be integrated in (\ref{genconv}), 
resulting in a $\chi_{m+1}$ with only absolute-value type kinks.

Below, we will show that  an arbitrary higher-derivative Lagrangian, 
(\ref{general}),  can be massaged so that 
all bulk terms have at most one $\partial_y$ 
acting on any given field, eg. $\chi^8 (\partial \chi)^{10}$,
 and all brane terms have no 
$\partial_y$'s  at all. As shown above, this will result in 
a well-defined perturbative derivative expansion for solving the equations 
of motion, with solutions having 
at most absolute-value type kinks.  The reader may on a first reading 
wish to accept that this can be done, skip the rest of this section, 
and continue 
with the stability analysis for higher-derivative bulk terms subject to the 
above conditions.

\subsection{Bulk higher-derivative operators}

 In this section 
we will show how higher-derivative operators in the bulk can be massaged 
using field redefinitions.
To simplify the considerations note   that 
the equations of motion subject to the 4D Poincare ansatz, $\chi = 
\chi(y)$, 
can always be obtained by first imposing the ansatz on the action, 
(\ref{general}), 
 and 
then functionally differentiating with respect to $\chi(y)$. 
Doing this we can write the bulk part of the Lagrangian in the form
\begin{equation}
{\cal L}_{bulk} =  M^3 \left\{ - \frac{1}{2}
 K(\chi) (\partial_y \chi)^2 - V(\chi) + 
\sum_{\vec{n}} \frac{a_{\vec{n}}(\chi)}{M^{|n|}}  (\partial \chi)^{n_1} ... 
(\partial^N \chi)^{n_N} \right\},
\end{equation}
where  $|n| \equiv n_1 + 2n_2 ... + N n_N - 2 > 0$, and where now 
$\partial \equiv \partial_y$. 

As already discussed, the problematic terms are those with $N > 1$
To begin, we assign the bulk potential a formal 
strength $V \sim {\cal O}(k^2) < M^2$ and augment our effective field theory 
derivative expansion with an expansion in $k/M$.
We then 
make a field redefinition, $\psi(\phi)$, to render the kinetic term 
``canonical'', 
\begin{equation}
\psi \equiv \int d \chi K^{1/2}(\chi).
\end{equation}
Then we have a form
\begin{equation}
{\cal L}_{bulk} = r_c M^3 \left\{ \frac{1}{2} (\partial \psi)^2 - V(\psi) + 
\sum_{\vec{n}} \frac{a_{\vec{n}}(\psi)}{M^{|n|}}  (\partial \psi)^{n_1} ... 
(\partial^N \psi)^{n_N} \right\},
\end{equation}
where of course $V$ and $a_n$ are redefined too. 
This particular field redefinition will simplify stating our procedure, and 
will be undone at the end.

We begin by working to zeroth order in $k/M$ (that is, we can neglect $V$), 
but to some non-zero order, $m$,  in $\partial/M$. Working inductively, 
we assume that by some field redefinitions of $\psi$ all $N >1$ terms 
have been eliminated from all terms of lower than $m$-th order. We then make 
the following field redefinition:
\begin{equation}
\psi \rightarrow \psi + \sum_{|n| = m} (-\partial)^{N-2} 
\frac{a_{\vec{n}}(\psi)}{M^{|n|}}  (\partial \psi)^{n_1} ... 
(\partial^N \psi)^{n_N-1}.
\end{equation}
Note that this transformation is only sensible for $N > 1$.
To $m$-th order, the only difference this makes to the Lagrangian 
arises from substituting into the zeroth order kinetic term, whereupon 
(after integrating by parts) it produces precisely the term needed to 
cancel $\sum_{|n| = m} \frac{a_{\vec{n}}(\psi)}{M^{|n|}}  (\partial 
\psi)^{n_1} ... (\partial^N \psi)^{n_N} $ in the Lagrangian, plus terms of 
higher than $m$-th order. Thus one uses this procedure to eliminate $N >1$ 
terms to any desired order in $\partial/M$, when we neglect ${\cal O}(k^2)$.  

More generally, 
the field transformation above
 will, however, reintroduce terms of various orders in $\partial/M$, but now 
with coefficients of 
 order $k^2/M^2$. But we can once again do a field redefinition at order 
$k^2/M^2$ and work to any desired order in $\partial/M$ to eliminate 
$N>1$ terms. This will in turn induce terms of order $k^4/M^4$. In this 
way,  one can eliminate all $N>1$ terms to any fixed order in 
$k^2/M^2$ and $\partial/M$.  Finally, we can 
transform back to a field, $\chi_{new}$: 
\begin{equation}
\frac{\partial \psi_{new}}{\partial \chi_{new}} = K^{1/2}(\chi_{new}), 
\end{equation} 
in terms of which, 
\begin{equation}
{\cal L}_{bulk} = r_c M^3 \left\{ \frac{1}{2}
 K(\chi_{new}) (\partial \chi_{new})^2 - V(\chi_{new}) + 
\sum_{n} \frac{a_{n}(\chi_{new})}{M^{n - 2}}  (\partial 
\chi_{new})^n \right\}.
\end{equation}
We have thereby removed all the problematic bulk terms discussed in the 
previous subsection.

It can be shown (again, tediously) that our field redefinition 
in Poincare ansatz,
\begin{equation}
\chi(y) \rightarrow \chi_{new}(y), 
\end{equation}
can be lifted to a fully 5D covariant transformation, 
\begin{equation}
\chi(x,y) \rightarrow \chi_{new}(x, y), 
\end{equation}
which, however, reduces to $\chi(y) \rightarrow \chi_{new}(y)$ upon imposing 
the 4D Poincare ansatz. Similarly, when gravity is included there are 
fully 5D generally covariant field redefinitions which upon imposing 
the Poincare ansatz ensure that all $N>1$ terms (terms with more than 
one $\partial_y$ acting on a field) are absent. 

\subsection{Brane higher-derivative operators}

It is obvious that the field transformations which we discussed above to 
massage the bulk action will induce derivative terms on the orbifold 
fixed points. Furthermore, such derivative terms may already be present 
anyway. Therefore after ensuring that all bulk terms satisfy $N \leq 1$, 
we must still massage away  brane terms in the Lagrangian 
of the form, 
\begin{equation}
{\cal L}_{brane} = \int d y \delta(y - y_i) 
M^3 k \left\{ \sum_{\vec{n}: N>0} \frac{a_{\vec{n}}(\chi)}{M^{|n|}}  (\partial 
\chi)^{n_1} ... 
(\partial^N \chi)^{n_N} \right\}.
\end{equation}
 We will show that the important physics contained 
in $\chi$  away from the branes can be obtained by using  
effective renormalized 
brane Lagrangians without any extra-dimensional derivatives,
\begin{equation} ~\label{braneeff}
{\cal L}_{eff.~brane} = \int d y \delta(y - y_i) 
M^3 k v_i(\chi).
\end{equation}
%A related discussion of classical brane renormalization appears in Ref. 
%\cite{wise}. Earlier work on classical renormalization appears in Ref. 
%\cite{damour}.  

The first step before renormalizing the original Lagrangian is to regulate 
it, say by the replacement,
\begin{eqnarray}
\delta(y - y_i) \rightarrow D(y - y_i) &\equiv& N 
e^{-1/(y - y_i + t)} e^{1/(y - y_i
 - t)}, ~~ - t < y < t \nonumber \\
&~& 0, ~~ {\rm else},
\end{eqnarray}
where 
\begin{equation}
N^{-1} = \int_{-t}^t d y e^{-1/(y + t)} e^{1/(y - t)}.
\end{equation}
This replaces the infinitesimally thin brane by a thick  brane 
of thickness $t$: $1/M < t << r_c$. Note that the brane 
profile $D(y - y_i)$
is a smooth function
(that is, a $C^{\infty}$ function) of compact support. 

With this regulator in place, clearly our perturbative derivative expansion 
is always well-defined despite derivatives on branes, and
(\ref{genconv}) will always give smooth solutions. 
Of course, the results depend on our 
choice of regulator. However we will show that away from the (compact) core 
of the branes, this 
regulator-dependence can be completely subsumed into an effective 
brane Lagrangian, (\ref{braneeff}). 

For simplicity of exposition we will consider a single orbifold fixed point on
${\bf R}/{\bf Z}_2$ rather than the two fixed points of $S^1/{\bf Z}_2$. The generalization to 
$S^1/{\bf Z}_2$ is entirely straightforward. First, 
recall that we have shown that 
at every order in perturbation theory we only solve second order 
differential equations. Thus the solution $\chi(y)$ in perturbation theory 
is completely specified by $\chi(0)$ and  $\partial_y \chi(0)$. Orbifold 
parity sets $\partial_y \chi(0) = 0$, but $\chi(0)$ (on ${\bf R}/{\bf Z}_2$ not $S^1/{\bf Z}_2$)
is an unfixed number, $\chi(0) = c$. In particular by solving the 
equations of motion, $\chi(t) = \chi(-t)$ and $\partial_y \chi(t) = - 
\partial_y \chi(-t)$ are both functions of $c$, 
\begin{eqnarray}
 \chi(t) &=& f(c) \nonumber \\
\partial_y \chi(t) &=& g(c).
\end{eqnarray}
Let us define $\tilde{\chi}(y)$, to be the result of integrating 
the second order differential equations from $t$ to a general $y > 0$, 
using $\chi(t), \partial_y \chi(t)$ as boundary conditions, 
 but completely neglecting brane interactions. 
Now for $y > t$, $\tilde{\chi}(y)$ is in fact the correct solution $\chi(y)$, 
since the neglected brane interactions vanish in this region. But 
for $y<t$, clearly $\tilde{\chi}(y)$ cannot be trusted. Nevertheless, 
as long as the physics of interest is dominated by bulk field behavior 
outside the core of the brane, we can use $\tilde{\chi}(y)$. This will be 
the case for the solution to the hierarchy problem, which is determined 
by the warp factor accumulated over the large bulk ($r_c \gg 1/M$).

Let us define, 
\begin{eqnarray}
\tilde{\chi}(0_+) &=& F(\chi(t), \partial_y \chi(t)) \nonumber \\
\partial_y \tilde{\chi}(0_+) &=& G(\chi(t), \partial_y \chi(t)).
\end{eqnarray}
We can eliminate dependence on $\chi(t), \partial_y \chi(t)$ in favor of $c$, 
\begin{eqnarray}
\tilde{\chi}(0_+) &=& F(f(c), g(c)) \equiv p(c) \nonumber \\
\partial_y \tilde{\chi}(0_+) &=& G(f(c), g(c)) \equiv q(c).
\end{eqnarray}
 Therefore, 
order by order in perturbation theory we can eliminate dependence on $c$
by inverting $p$, 
\begin{equation}~\label{chitilde}
\partial_y \tilde{\chi}(0_+) = q(p^{-1}(\tilde{\chi}(0_+))).
\end{equation}

We will now show that $\tilde{\chi}$ is the classical solution corresponding 
to a ``renormalized'' effective Lagrangian with the same bulk terms as before,
but with a $\delta$-function brane-localized potential term (that is without 
any extra-dimensional derivatives). Recall that by field redefinitions we 
have already ensured that the bulk Lagrangian only depends on $\chi$ and 
its first derivative, ${\cal L}_{bulk}(\chi, \chi')$. The full 
effective Lagrangian, including the effective brane potential, is then given 
by, 
\begin{equation}~\label{effder}
{\cal L}_{eff} = {\cal L}_{bulk}(\chi, \chi') + \delta(y) v_{eff}(\chi),
\end{equation}
where,
\begin{equation}
v_{eff}(\chi) = \int d \chi ~2 q(p^{-1}(\chi)) \frac{\partial^2 
{\cal L}_{bulk}}{(\partial \chi')^2}\left(\chi, q(p^{-1}(\chi))\right).
\end{equation}
The reader can straightforwardly check that the equation of motion that 
follows from this effective Lagrangian is equivalent to 
 the equation of motion 
due only to the bulk terms away from $y=0$, supplemented by the boundary 
condition, (\ref{chitilde}). 

Thus, we are always able to find an effective Lagrangian of the form
(\ref{effder}), which has solutions which agree with those of a general 
Lagrangian supplemented by a regulator, 
outside the regulated core of the 
branes. The important physics such as the hierarchy are insensitive to the 
general disagreement inside the thickness of the regulated brane.
Note  that the effective 
Lagrangian does not suffer from UV ambiguities, and therefore needs no 
regulator. In that sense it is ``renormalized''. 

\section{Stability and Higher Derivative Operators}

\subsection{Set-up of the model}

We will now show that the  Goldberger-Wise mechanism 
can be realized in a very general setting, including higher-derivative 
operators. 
We will take the higher-derivative terms
 to be suppressed by appropriate powers 
of $1/M$ and constrained by symmetries according to the discussion of 
Section III\@. Furthermore, we will assume that they have already been 
massaged by field redefinitions and brane-renormalization as discussed in 
the previous section, so that fields appearing in bulk interactions 
have at most one extra-dimensional derivative acting on them, while there 
are no extra-dimensional derivatives in brane terms.

 The action in Einstein 
frame  is 
\begin{eqnarray}
S_{bulk} & = &  M^3 \int d^4 x \int_{-\pi}^{\pi} d \phi \sqrt{G} 
\left( k^2 V(\chi) - \frac{1}{4} R + \frac{1}{2} K(\chi) (D \chi)^2  
+ M^2 {\cal L}_{h.d.}(G_{MN}, \chi, \frac{\partial_N}{M}) \right) \\
S_{vis} & = & - M^3 k \int d^4 x \sqrt{g_{vis}} v_{v}(\chi)  \\
S_{hid} & = & - M^3 k \int d^4 x \sqrt{g_{hid}} v_{h}(\chi).
\end{eqnarray}
Here, ${\cal L}_{h.d.}$ indicates terms containing more than two derivatives
with dimensionless coefficients of order unity 
and suppressed by powers of $1/M$.
We can expand the dimensionless functions $V, K, v_v,$ and 
$v_h$ in powers of $\chi$.
\begin{eqnarray}
V(\chi) & = & 3  - \frac{1}{2} \epsilon \chi^2 + {\cal O}(\chi^3) \nonumber \\
K(\chi) & = & 1 + {\cal O} (\chi) \nonumber \\
v_v(\chi) & = & \rho_{v} + \lambda_{v} \chi + \frac{1}{2} \mu_v \chi^2 + {\cal O}(\chi^3) \nonumber \\
v_h(\chi) & = & \rho_{h} + \lambda_{h} \chi + \frac{1}{2} \mu_h \chi^2 + {\cal O}(\chi^3).
\end{eqnarray}

We will perform perturbation theory in powers of $k^2/M^2 \ll 1$ and 
the brane-tadpoles, $\lambda_{v,h}$, as discussed in Section II\@. 
Higher-derivative terms are automatically suppressed by powers of $k^2/M^2$ 
because the bulk potential is dominated by a 5D cosmological constant of 
order $M^3 k^2$. Formally, this is seen by the fact that in our 
dimensionless $y$-coordinate every $\partial_y$ is accompanied by 
$k$. For 
simplicity,
 we take $k^2/M^2$ and $\lambda_{v,h}$ to all be of the order of our 
formal small parameter, $\lambda$. We continue to take the relations of 
(\ref{epsiloncondition}) to hold among $\lambda_{v,h}, 
\rho_{v,h}, \mu_{v,h}$ and $\epsilon$.

We now write the equations of motion with the Poincare ansatz using again the dimensionless variable
$y = k r_c \phi$ as in Section II, 
\begin{eqnarray}
\chi'' + 4 A' \chi' - \epsilon \chi & = & (\lambda_{v} +\mu_v \chi + {\cal O}(\chi^2)) 
\delta (y - y_c) + (\lambda_{h} + \mu_h \chi + {\cal O}(\chi^2)) \delta (y) \nonumber \\ 
& & \mbox{} \qquad + f_1(\chi, \chi', A') + \chi'' f_2(\chi, \chi', A') \label{d} \\
A'' + \frac{2}{3} (\chi')^2 & = & -\frac{2}{3} (\rho_{v} + \lambda_{v} \chi + 
\frac{1}{2} \mu_v \chi^2 +{\cal O}(\chi^3)) \delta(y-y_c) - \frac{2}{3} (\rho_h +  \lambda_{h}
 \chi +\frac{1}{2} \mu_h \chi^2 + {\cal O}(\chi^3)) \delta(y) \nonumber \\
& & \mbox{} \qquad 
+ f_3(\chi, \chi', A') + A'' f_4(\chi, \chi', A')  \label{e} \\
(A')^2 & = & 1 - \frac{1}{6} \epsilon \chi^2 + \frac{1}{6} (\chi')^2 + 
f_5(\chi, \chi',A'), \label{f}
\end{eqnarray}
where the $f_i$ refer to the variation of the bulk higher-derivative action 
subject to the special form discussed in Section IV\@.  
Note that the right-hand side of  (\ref{f}) does not contain terms 
with delta functions or second derivatives.

Perturbation theory follows along lines similar to  
Section II\@. We first formally expand $A$ and $\chi$ as in (\ref{expansion}) 
and substitute into the equations of motion. We then solve (\ref{f}) for the 
$A'_n$ in terms of the $\chi_{m < n}$ and  $\chi'_{m < n}$ 
to any desired order. Substituting 
for the $A'_n$ into  (\ref{d}) then yields an equation involving only the 
$\chi_n$ and $\chi'_n$ which we solve perturbatively. Then, by 5D 
general covariance in the bulk, (\ref{e}) is automatically solved, except 
for the two brane junction conditions which are solved by 
fine-tuning $\rho_h$ and setting $y_c$ to its stable vacuum value.  

 To see  that (\ref{e}) is automatically satisfied up to junction conditions, 
note that
 under infinitesimal general coordinate transformations the bulk action is invariant,
\begin{equation}
S_{bulk} \left[ G_{MN}+D_M \eta_N + D_N \eta_M, \chi + \partial^M \chi \eta_M \right] = 
S_{bulk} \left[ G_{MN},\chi \right],
\end{equation}
for arbitrary infinitesimal $\eta_M$.  This implies
\begin{equation}
2 D_M \left( \frac{1}{\sqrt{G}} \frac{\delta S_{bulk}}{\delta 
G_{MN}} \right) - \frac{1}{\sqrt{G}} \frac{\delta S_{bulk}}{\delta 
\chi} \partial^N \chi = 0.
\end{equation}
In the Poincare ansatz this gives
\begin{equation}
\partial_5 \left( \frac{\delta S_{bulk}}{\delta G_{55}} \right) = 
A' \frac{\delta S_{bulk}}{\delta G^{\mu}_{\mu}} - 
\frac{1}{2} \chi' \frac{\delta S_{bulk}}{\delta \chi}.
\end{equation}
This shows that  (\ref{e}), the $G_{\mu \nu}$ equation of motion, is 
satisfied
up to junction conditions, given (\ref{d}), 
 the $\chi$ equation of motion, and (\ref{f}), 
the $G_{55}$ equation of motion.  

The equation for $\chi$ obtained by eliminating $A'$ from (\ref{d}) using 
(\ref{f}) has the form (\ref{recurrence}), but now with sources following 
from (\ref{d}-\ref{f}) of the form
\begin{equation}~\label{newJ}
J_n = J^{bulk}_n + J^{brane}_n + J^{mixed}_n, 
\end{equation}
where 
\begin{equation}
J_{n}^{bulk} = \sum a_{r,s,\vec{i}} \lambda^{m} \chi_{i_1} \ldots \chi_{i_{r}} \chi'_{i_{r+1}} \ldots \chi'_{i_s}  \delta_{i_1 + \cdots + i_s + m,~ n}
\end{equation}
\begin{equation}
J_{n}^{brane_j} = \sum b_{r,s,\vec{i}} \lambda^m \chi_{i_1} \ldots \chi_{i_{s}} \delta_{i_1 + \cdots i_s + m, n} \delta(y - y_j),
\end{equation}
and 
\begin{equation}
J_{n}^{mixed} = \sum c_{r,s,\vec{i}} \lambda^m \chi_{i_1} \ldots \chi_{i_{s}} \chi'_{i_{r+1}} \ldots \chi'_{i_s}  \chi''_p \delta_{i_1 + \cdots i_s + 
m + p, ~ n}.
\end{equation}
This last term acts as a brane term as well as a bulk term because $\chi''$ 
contains  $\delta$-functions at the branes as well as a smooth bulk behavior.
Again, the constant coefficients, $a, b$ and  $c$ are order one or smaller.
The $n^{th}$ order solution is obtained using (\ref{convolution}) with the 
same Green function defined by (\ref{Gdef}).

\subsection{Stabilization at second order}

We will show here that hierarchy stabilization satisfying (\ref{claim}) 
is naturally achieved by second order in $\lambda$, similarly to Section II\@.
Note that at zeroth order our solution is again the unstabilized RS1 
solution. At first order $\chi_1$ is given again by (\ref{chi1}), although 
$A_1'$ may now be a non-zero constant. As discussed 
above, we can solve for the $A_n'$ using (\ref{f}) in terms of the $\chi_{m <n}$ and  
$\chi'_{m <n}$. Therefore, $A_2'$ must be a quadratic 
polynomial in $\chi_1$ and $\chi_1'$. Summarizing, we have
\begin{equation}~\label{alpha}
A_0' + A_1' + A_2' = \alpha_0 + \alpha_1 \chi_1 + \alpha_2 \chi_1'
+ \frac{1}{12} \epsilon \chi_1^2 - \frac{1}{12} \chi_1^{\prime 2},
\end{equation}
where $\alpha_0 = - 1 + {\cal O}(\lambda), ~ \alpha_{1,2} 
\sim {\cal O}(\lambda)$. The constants 
$\alpha_i$ are independent of $y_c$ since the $A'_n$ are determined in terms 
of the $\chi_m$ and  $\chi'_m$ locally.

It remains to satisfy the junction conditions for (\ref{e}).
After an ${\cal O}(\lambda^2)$ tuning of $\rho_v$
 (not appearing 
at higher orders), similar to Section II, the above
considerations and (\ref{chi1})  yield a visible junction condition
 of the form
\begin{equation}
P_2(e^{\Delta_- y_c}) = 0,
\end{equation}
where $P_2$ is a quadratic polynomial with 
order one coefficients.  For generic values of these coefficients 
one obtains solutions,
\begin{equation}
e^{\Delta_- y_c} = {\cal O}(1).
\end{equation}
Thus, along with (\ref{alpha}) we have demonstrated the hierarchy, 
(\ref{claim}), at this order. 
The hidden brane junction condition is solved by fine-tuning $\rho_h$, 
corresponding to the fine-tuning of the 4D cosmological constant to zero.

\subsection{Subdominance of higher orders}

We claim that at any order $\chi_n$ has the general form (\ref{chin}), 
and $A'_n$ has the general form, 
\begin{equation}~\label{aprime}
A'_n =  \sum q_{\vec{n},r,s, t} \lambda^{n_1} (\lambda y)^{n_2} 
( \lambda y_c)^{n_3} (\lambda/\Delta_-)^{n_4} e^{r \Delta_+ (y-y_c)} 
e^{s \Delta_- y} e^{t \Delta_- y_c}.
\end{equation}
so that perturbation theory is well-behaved. Therefore, our second order 
mechanism above for stabilization of the weak/Planck hierarchy of order 
$e^{-{\cal O}(1)/\epsilon}$ continues to hold.

%\begin{eqnarray}
%\chi_n & = & \sum d_{\vec{n}, r,s} \lambda^{n_1} (\lambda y)^{n_2} 
%( \lambda y_c)^{n_3} (\lambda/\Delta_-)^{n_4} e^{s \Delta_+ (y-y_c)} 
%e^{t \Delta_- y}, 
%%A'_n & = &  \sum b_{\vec{n}, r,s} \lambda^{n_1} 
%(\lambda y)^{n_2} ( \lambda y_c)^{n_3} (\lambda/\Delta_-)^{n_4} 
%e^{r \Delta_+ (y-y_c)} e^{t \Delta_- y}
%\end{eqnarray}
%with order one or smaller coefficients, $d$.  The integers $n_i, r, s$ 
%satisfy
%\begin{eqnarray}
%n_1 & \ge & 1; r,n_2,n_3,n_4 \ge 0 \\
%n & = & n_1 + n_2 + n_3 + n_4 \\
%|s| & < & n.
%\end{eqnarray}
%We will show that the  Weak/Planck hierarchy is then 
%set by $e^{-{\cal O}(1)/\epsilon}$.  

We show this claim is true by induction.  We have already discussed the 
cases for $n=0,1$. Focusing now on $\chi_n$ and 
assuming that (\ref{chin}) is valid at all orders lower than $n$, the 
$n^{th}$ order source has the form
\begin{eqnarray}
J_n & = & \sum a_{\vec{n},r,s, t} \lambda^{n_1} (\lambda y)^{n_2} ( \lambda y_c)^{n_3} (\lambda/\Delta_-)^{n_4} e^{r \Delta_+ (y-y_c)} 
e^{s \Delta_- y} e^{t \Delta_- y_c} \nonumber \\
 &  & \mbox{} + \sum b_{\vec{n},r,s, t} \lambda^{n_1} (\lambda y)^{n_2} ( \lambda y_c)^{n_3} (\lambda/\Delta_-)^{n_4} e^{- r \Delta_+ y_c} e^{t \Delta_- y_c} 
\delta(y)  \nonumber  \\
& & \mbox{} + \sum c_{\vec{n},r,s, t} 
\lambda^{n_1} (\lambda y)^{n_2} ( \lambda y_c)^{n_3} (\lambda/\Delta_-)^{n_4} e^{- r \Delta_+ y_c} e^{t \Delta_- y_c} 
\delta(y-y_c).
\end{eqnarray}
It is straightforward to check that with this source, the solution for 
$\chi_n$ given by (\ref{convolution}) indeed satisfies (\ref{chin}). 
Then, given  the form of (\ref{f}) and the form for $\chi_n$ of (\ref{chin}),
 (\ref{aprime}) readily follows.

\subsection{The case of $\lambda > \epsilon$}

A significant difference between the source considered in this section, 
(\ref{newJ})  and 
that  considered in Section II, (\ref{rootexp}),  is that in the latter 
every power of $\chi$ without a derivative was accompanied by an explicit 
power of $\sqrt{\epsilon}$. Recall that this ensured that increasing powers 
of $1/\Delta_-$ in perturbation theory were accompanied by $\epsilon$ 
factors.    This allowed us to take $\lambda > \epsilon$ and 
still have a meaningful perturbative expansion. Without this feature, in this
section we are limited to formally 
$\lambda < \epsilon$, as we have assumed thus far in this section, 
in order to have a good 
expansion. 

However, there is an interesting scenario under which we would recover 
a good expansion for $\lambda > \epsilon$ even for higher-derivative 
operators. We first assume a (non-linearly realized) global 
symmetry\footnote{We thank W. Goldberger for pointing out the usefulness 
of this symmetry to us.},
\begin{equation}
\chi \rightarrow \chi + {\rm constant}. 
\end{equation}
If this symmetry were exact, only derivatives of $\chi$ would be 
permitted in the action.
But one does not expect that such global symmetries are respected by the 
underlying quantum gravity theory. We will assume that in fact such 
violation is controlled by the small parameter, $\epsilon$, so that 
non-derivative powers of $\chi$ in the action are accompanied by 
$\sqrt{\epsilon}$, thereby generalizing the explicit appearance of 
$\epsilon$ in Section II\@. Then, increasing powers of 
$1/\Delta_-$ in perturbation theory are again accompanied by $\epsilon$ 
factors, so that the perturbative expansion is under control for 
$\lambda > \epsilon$.

\section{Conclusions}

We demonstrated the stability of the RS mechanism for 
generating the weak/Planck hierarchy in the presence of  higher-derivative
interactions.
 We re-worked the Goldberger-Wise radius stabilization mechanism 
in a systematic perturbative expansion in parameters of the brane potential. 
Incorporating higher-derivative 
interactions as further perturbations, we showed that they did not 
affect the basic mechanism. 

In our perturbative 
analysis, radius stabilization is achieved in the following 
steps. At zeroth order we found the unstabilized RS1 vacuum 
with a trivial profile for the Goldberger-Wise scalar. At first order, the 
scalar responds non-trivially to the brane-potentials. The back-reaction 
of this scalar profile on the 5D metric (warp factor) takes place at 
second order.  In particular, the radius acquires a distinct, stable value. 
This value can naturally be several times the fundamental scale of the 
theory, while the bulk geometry is a small perturbation of AdS$_5$, so 
that the 
RS mechanism for generating the hierarchy operates.
We also gave a careful treatment of all higher orders in perturbation 
theory, showing that our expansion is under good 
control.  

The 
derivative expansion of effective field theory seems at first fundamentally 
at odds with the presence of ``thin'' or $\delta$-function 
branes or orbifold fixed points, resulting in ill-defined 
singularities in the classical equations of motion. While such 
singularities can be regulated, the process is messy and apparently 
 regulator-dependent. However, 
we showed how classical renormalization of the
brane action can be implemented so as to remove the need for explicit
regulators. This greatly simplifies
 the discussion of higher-derivative effects.

It is important to study the stability of the RS1 effective 
field theory under quantum effects. Since the effective theory is 
not renormalizable we expect quantum divergences of the form of every 
possible local operator (subject only to symmetries). 
However, once these 
divergences are covariantly regulated at the fundamental scale, 
they must be of the form and strength  
of the higher-derivative interactions
 already considered in this paper. As usual in 
effective field theory, 
renormalization proceeds order by order in the derivative expansion, 
using counterterms also of the form of the operators of this paper. 
Thus, because quantum divergences are local they cannot destabilize the RS
hierarchy, since our purely 
classical analysis already treats all such local effects. This leaves 
only the non-local quantum effects which are UV-finite and therefore 
well-defined and calculable. In future work, we hope to build on Refs. 
\cite{rothstein} in studying the general structure of these non-local  
quantum amplitudes and whether they affect the basic RS1 mechanism.

\acknowledgements

We are grateful to Walter Goldberger, Markus Luty, Rustem 
Ospanov, Alexey Petrov and Mark Wise for helpful discussions. 
A.L. was supported in part by the National Science Foundation under 
grant PHY-9970781.

\end{document}